\documentclass[pra,twocolumn,preprintnumbers,amsmath,amssymb,superscriptaddress,
  10pt]{revtex4}
\usepackage{graphicx}
\usepackage{bm}
\usepackage{color}
\usepackage{notes2bib}
\DeclareMathOperator{\e}{e}

\def\Xint#1{\mathchoice
   {\XXint\displaystyle\textstyle{#1}}%
   {\XXint\textstyle\scriptstyle{#1}}%
   {\XXint\scriptstyle\scriptscriptstyle{#1}}%
   {\XXint\scriptscriptstyle\scriptscriptstyle{#1}}%
   \!\int}
\def\XXint#1#2#3{{\setbox0=\hbox{$#1{#2#3}{\int}$}
     \vcenter{\hbox{$#2#3$}}\kern-.5\wd0}}

\def\dashint{\Xint-}

\begin{document}

\title{Transparency enhancement in a double-barrier structure by the Fano antiresonance}

\author{J. Klier}
\affiliation{Institut f\"ur Nanotechnologie, Karlsruhe Institute of Technology, 76021 Karlsruhe, Germany}

\author{I.V. Krainov}
\affiliation{Ioffe Physico-Technical
Institute of the Russian Academy of Sciences, 194021 St.~Petersburg,
Russia} 
\affiliation{Lappeenranta University of Technology, P.O. Box 20,
FI-53851, Lappeenranta, Finland}

\author{A.P. Dmitriev}
\affiliation{Ioffe Physico-Technical
Institute of the Russian Academy of Sciences, 194021 St.~Petersburg,
Russia} 

\author{I.V. Gornyi}
\affiliation{\mbox{Institute for Quantum Materials and Technologies, Karlsruhe Institute of Technology, 76021 Karlsruhe, Germany}}
\affiliation{Ioffe Physico-Technical
Institute of the Russian Academy of Sciences, 194021 St.~Petersburg,
Russia} 


\date{\today}

\begin{abstract}
We show that the presence of a side-attached state strongly modifies the transmission through a one-dimensional double-barrier system in the window of wavevectors around the Fano antiresonance. 
Specifically, the interplay between the Fano interference and the size quantization inside the structure gives rise to narrow resonant peaks in the transmission coefficient. The height of the peaks may become close to unity (perfect transmission) even for an asymmetric setup with strong barriers, where the transmission coefficient in the absence of the Fano state is strongly suppressed at all other wavevectors. Thus, the two types of the interference phenomena, each by itself leading to the suppression of the transmission, conspire in a peculiar way to produce the transparency enhancement.   
\end{abstract} 

\maketitle

\section{Introduction}

Fano resonant suppression \cite{Fano} of the wave transmission in the presence of a localized discrete state is one of the most famous interference phenomena in atomic physics, optics, and electronics (for reviews, see Ref. \cite{FanoRMP,Book}). It is well known that the amplitude $A$ of the electron transition to the states of a continuum through a localized quasistationary level $\varepsilon_0$ is described by the Breit-Wigner formula 
\begin{equation}
A=iA_0\frac{\Gamma}{\varepsilon-\varepsilon_0+i \Gamma},
\end{equation}
where $\varepsilon$ is electron's energy, $\Gamma$ is the width of the level, and the magnitude $A_0$ does not vary with energy on the scale of the order of $\Gamma$. The probability of transition demonstrates the symmetrical resonance of the usual Lorentz form.  If, in addition to the resonant transition through the intermediate level, a direct transition with an amplitude $B$ that weakly depends on energy is also allowed, the total transition amplitude is equal to the sum of the amplitudes  $A$ and $B$. Then an interference term $2\text{Re}(A B^*)$  appears in the expression for the transition probability
\begin{align}
W
=|A+B|^2
=|B|^2\frac{|\varepsilon-\varepsilon_0+i\Gamma(A_0/B+1)|^2}
{(\varepsilon-\varepsilon_0)^2+\Gamma^2},
\end{align} 
which leads to an asymmetric resonance, first described by Fano \cite{Fano} for the case of photoionization of the atom. When $A_0/B+1=i q$ is purely imaginary, the asymmetry of the Fano resonance is governed by $q$,
 \begin{align}
W
=|B|^2\frac{(\varepsilon-\varepsilon_0-q\Gamma)^2}
{(\varepsilon-\varepsilon_0)^2+\Gamma^2},
\label{Fano-q}
\end{align}
which interpolates between the Breit-Wigner resonance for large $q$ and the symmetric Fano antiresonance at $q=0$. In the latter case, the transition probability is strictly zero at $\varepsilon=\varepsilon_0$.

Since then, a wide variety of physical systems, from mechanical to nuclear, have been studied experimentally and theoretically, in which one of the two interfering amplitudes for transitions into the continuum is resonant. In particular, such a situation is realized for electrons passing through a one-dimensional (1D) system, next to which a tunnel-coupled ``atom'' is located \cite{FanoRMP,JETP}. An electron trapped on the atom's level has the energy $\varepsilon_0$ and the lifetime $\hbar/\Gamma$ corresponding to the tunneling coupling between the atom and the wire. The transmission coefficient is then proportional to the ratio $(\varepsilon-\varepsilon_0)^2/[(\varepsilon-\varepsilon_0)^2+\Gamma^2]$, which corresponds to $q=0$ in Eq.~(\ref{Fano-q}).  This means that at $\varepsilon=\varepsilon_0$, the wave corresponding to the capture of an electron on the atom completely ``extinguishes'' the directly transmitted wave, so that there is full reflection (Fano antiresonance in transmission). 

Physics related to the Fano antiresonance is relevant, for example, to complex quantum-dot structures \cite{Weijiang}, such as a tunnel-coupled carbon nanotubes (CNT) with side-attached single-molecule magnets \cite{GMRMain,zener,GMRCNT,TbPc2CNTDFTKorea}. These structures received much attention because of the giant magnetoresistance which is caused by a spin-dependent scattering of the conducting electrons on the localized state of a single-molecule magnet~\cite{GMRMain,Krainov} (see also Ref.~\cite{Titov} for a related discussion of spin filter with Fano states). The calculations in the present paper are motivated by our recent work \cite{Krainov} that proposed an explanation of the giant magnetoresistance observed in the Coulomb blockade regime in CNTs with with organic molecules attached to them. The molecules create a quasistationary discrete level for electrons of the CNT, leading to the Fano resonance in the transmission coefficient. In Ref.~\cite{Krainov}, in accordance with the conditions of the experiments, a rather fast phase breaking was assumed, so that the electron motion between two tunnel barriers was not quantized (the Fano state on the molecule is tunnel-coupled to the continuous spectrum in the CNT). However, of independent interest is the problem of the effect of level quantization on the Fano antiresonance in a \textit{coherent} quantum dot. This paper is dedicated to solving this problem. 

In a 1D system with two strong barriers without a resonant impurity, the transmission coefficient as a function of the electron energy is a ``comb'' of narrow peaks centered at energies that coincide with the energies of the size quantization levels. We show that the presence of a side-attached state strongly modifies the transmission through a double-barrier system in the window of wavevectors around the Fano antiresonance. Specifically, the interplay between the Fano interference and the size quantization inside the structure gives rise to narrow resonant peaks in the transmission coefficient. Remarkably, the height of the peaks may become close to unity (perfect transmission) even for an asymmetric setup with strong barriers, where the transmission coefficient is strongly suppressed at all other wavevectors. 

Thus, the two types of the interference phenomena, each by itself leading to the suppression of the transmission, conspire in a peculiar way to produce the transparency enhancement. This striking phenomenon can be potentially used as a narrow-band filter in nanoelectronic and photonic devices. While engineering Fano resonances in complex structures has been a subject of a number of works (see, e.g., Ref.~\cite{Kivshar} and reviews\cite{FanoRMP,Book}), to the best of our knowledge, the emergence of a Fano-state-induced narrow resonances with almost perfect transmission in a double-barrier structure has not been previously appreciated in the literature.

The paper consists of the introduction, three main sections, and the conclusion. In Sec.~\ref{Sec:Basics}, we recapitulate the basics of the two main ingredients of our analysis separately and introduce the notations. First, the transmission of electrons through a 1D system with Fano resonance and without tunneling barriers is addressed in Sec.~\ref{Sec:Fano}. Next, a double-barrier system is considered in Sec.~\ref{Sec:Barriers}. In Sec.~\ref{Sec:Double}, the problem of the transmission of an electron through a 1D system with two tunnel barriers and Fano resonance between them is solved. In Sec.~\ref{Sec:Resonance}, the emergent narrow transmission resonance is analyzed in detail, first for identical barriers and then for a generic double-barrier structure. Technical details of an alternative derivation are relegated to Appendix.

\section{Basics}
\label{Sec:Basics}

\subsection{Transmission across a Fano state}
\label{Sec:Fano}

\begin{figure}
  \centering
  \includegraphics[width=\linewidth]{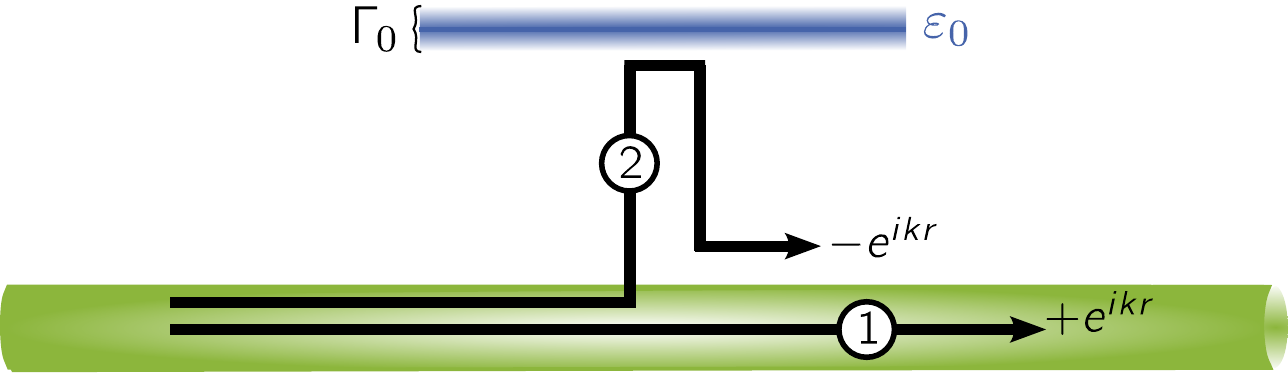}
  \caption[Schematic illustration of the origin of the Fano resonance in transmission though an infinite one-dimensional channel]{Schematic illustration of the origin of the Fano resonance in transmission though an infinite 1D channel tunnel-coupled with a discrete state (red level). 
The destructive interference of waves passing the discrete level without visiting it (path 1) and the waves visiting the Fano state (path 2) leads to the vanishing of the transmission coefficient for electrons with the resonant energy ($E_k = \varepsilon_0$).}
  \label{fig:Fano}
\end{figure}

We start with a brief reminder of the formalism employed to describe the influence of a Fano state on transport in an infinite one-dimensional (1D) channel \cite{Fano,FanoRMP}, see Fig.~\ref{fig:Fano}. Throughout the paper, we consider a single-particle problem. In the presence of electron-electron or electron-phonon interactions in a single-channel wire, the wire conductance in the presence of a side-attached localized state have been studied in Refs.~\cite{PhysRevLett.100.256805,PhysRevLett.104.106403} and \cite{Galda}.

The Hamiltonian of 1D free electrons which can tunnel to a side-attached localized state reads:
\begin{align}
\hat{\mathcal{H}} &= \sum_k E_k |\psi_k\rangle \langle \psi_k|\notag 
\\
 &+
\sum_k \big(
\varepsilon_0|\varphi\rangle \langle \varphi|+V_k |\varphi\rangle \langle \psi_k| + V_k^* |\psi_k\rangle \langle \varphi| \big).
\label{hFano}
\end{align}
Here, 
$\varepsilon_0$ denotes the energy of the localized state with the corresponding wave function 
$\varphi$, $E_k = \hbar^2 k^2/(2 m)$ and $\psi_k$ are the energy and wave function of state $k$ in the 1D channel, $V_k = \langle \varphi | \hat{V} | \psi_k \rangle$ is the Bardeen tunneling matrix element~\cite{Bardeen} of the tunneling operator $\hat{V}$.
The wave functions $\varphi$ and $\psi_k$ are assumed to be orthogonal to each other.

The transmission ($t_F$) and reflection ($r_F$) amplitudes 
for the localized Fano state are given by \cite{FanoRMP}:
\begin{eqnarray}
t_F &=& \frac{E_k - \varepsilon_0(k)}{E_k - \varepsilon_0(k) + i \Gamma_0(k)},
\label{trFano}
\\
r_F &=& \frac{-i \Gamma_0(k)}{E_k - \varepsilon_0(k) + i \Gamma_0(k)},
\label{refFano}
\end{eqnarray}
where (the dash denotes the principal-value integral)
\begin{gather}
\varepsilon_0(k) = \varepsilon_0 + \dashint_0^\infty dp\ \frac{|V_p|^2}{E_k - E_p}, 
\label{dE}
\\
\Gamma_0(k) = \pi \int_0^\infty dp\ \delta(E_p - E_k) |V_p|^2.
\label{Gamma0K}
\end{gather}
For a pointlike tunneling, $V_p$ is momentum independent and, hence, 
there is no $k$ dependent shift of the Fano-state energy and the suppression of the transmission is symmetric. In the general case,  we introduce $k_0 = \sqrt{2 m \varepsilon_0/\hbar^2}$ and $\Gamma_0=\Gamma_0(k_0)$ [for weak tunneling, it is sufficient to take $\varepsilon_0(k)$ and $\Gamma_0(k)$ at the resonance point]. The energy scale $\Gamma_0$ describes the broadening of the localized state, which stems from the coupling between the 1D channel and localized state. The Fano state suppresses the transmission in the window of resonant energies $E_k \sim (\varepsilon_0 - \Gamma_0, \varepsilon_0 + \Gamma_0)$. For a confined geometry, however, the broadening of the Fano resonance is changed by the properties of the boundaries and may also be momentum dependent, as we will discuss in the following sections.

A schematic illustration of the processes responsible for the Fano resonance is presented in Fig.~\ref{fig:Fano}. This figure shows the interference of the electronic wave that passes through the wire with the continuous spectrum directly, i.e., without visiting the Fano state, and the wave that passes via scattering on the discrete state (paths 1 and 2 in Fig.~\ref{fig:Fano}, respectively). Exactly at the resonance, $E_k = \varepsilon_0$, path 2 acquires a phase of $\pi$ leading to zero transmission, since the sum of the two waves vanishes $e^{ikx}+e^{ikx+i\pi}=0$.
Accordingly, the reflection coefficient reaches maximum (unity) at resonance. 

\subsection{Double-barrier structure}
\label{Sec:Barriers}

In this section, we present a description of a double-barrier structure in the absence of the Fano state.
Again, as in Sec.~\ref{Sec:Fano} above, we do not consider here effects of electron-electron 
interactions in the wire, which would lead to a peculiar renormalization of the transmission coefficient through the structure, see Refs.~\cite{KF,FN,NG,PG}.

Without loss of generality, we model the tunnel contacts as two $\delta$-function barriers 
located at $x=\pm L$,
with the potentials
\begin{equation}
V_{L,R}(x)=\eta_{L,R} \frac{\hbar^2}{2m}\delta(x \pm L),
\end{equation} 
characterized by the strengths $\eta_{L}$ and $\eta_{R}$ for the left and right barrier, respectively. 
The transmission and reflection amplitudes for a single delta-barrier
are given by
\begin{equation}
t_B=\frac{2ik}{2ik-\eta}, \quad r_B=\frac{\eta}{2ik-\eta}.
\end{equation}

We introduce the wave-functions for the system with these two barriers  
and decompose it into the symmetric and antisymmetric parts as
\begin{align}
&\Psi_{k+}(x)\!=\!\left\lbrace\begin{array}{ll}
\cos(kx)+\frac{\eta_R \cos(kL)}{k}\sin[k(x-L)], &x\geq L,\\
\cos(kx), &|x|\leq L,\\
\cos(kx)-\frac{\eta_L \cos(kL)}{k}\sin[k(x+L)], &x\leq -L,
\end{array}\right.
\end{align}
\begin{align}
&\Psi_{k-}(x)\!=\!\left\lbrace\begin{array}{ll}
\sin(kx)+\frac{\eta_R \sin(kL)}{k}\sin[k(x-L)], &x\geq L,\\
\sin(kx),  &|x|\leq L,\\
\sin(kx)+\frac{\eta_L \sin(kL)}{k}\sin[k(x+L)], &x\leq -L.
\end{array}\right.
\end{align}
The right-moving wave is described by
\begin{align}\label{wf_r}
\phi_{+k}(x)=a_{r+}(k)\Psi_{k+}(x)+a_{r-}(k)\Psi_{k-}(x),
\end{align}
and the left moving wave is given by 
\begin{align}\label{wf_L}
\phi_{-k}(x)=a_{l+}(k)\Psi_{k+}(x)-a_{l-}(k)\Psi_{k-}(x).
\end{align}
Here, the coefficients are given by
\begin{align}\label{aR+}
a_{r+,l+}(k)&=\frac{2ik}{D(k)}\left[k+\eta_{R,L}\sin(kL)\exp(ikL)\right], 
\\
\label{aR-} 
a_{r-,l-}(k)&=\frac{2ik}{D(k)}\left[ik-\eta_{R,L}\cos(kL)\exp(ikL)\right],
\end{align}
with
\begin{gather}
	D(k)=[ik-\eta_L\cos(kL)e^{ikL}][k+\eta_R\sin(kL)e^{ikL}] \notag
	\\
	+[k+\eta_L\sin(kL)e^{ikL}][ik-\eta_R\cos(kL)e^{ikL}]. \label{DK}
\end{gather}

For the case of symmetric setup with $\eta_R=\eta_L=\eta$, 
we get $a_{r+}=a_{l+}=a_+$ and $a_{r-}=a_{l-}=a_-$, where
\begin{align}\label{a+}
a_+(k)&=\frac{ik}{ik-\eta\cos(kL)\exp(ikL)}, \\
\label{a-} a_-(k)&=\frac{ik}{k+\eta\sin(kL)\exp(ikL)}.
\end{align}
Using these equations, we express the transmission and reflection coefficients of the symmetric 
double-barrier structure as:
\begin{align}\label{TBB}
t_\text{BB}(k)&=\frac{1}{2}\left[\frac{a_+(k)}{a_+^{\ast}(k)}-\frac{a_-(k)}{a_-^{\ast}(k)}\right],\\
\label{RBB}
r_\text{BB}(k)&=\frac{1}{2}\left[\frac{a_+(k)}{a_+^{\ast}(k)}+\frac{a_-(k)}{a_-^{\ast}(k)}\right].
\end{align}
These formulas describe resonances at energies corresponding to the size quantization levels.
For the symmetric setup, the transmission coefficient 
\begin{equation}
\mathcal{T}_\text{BB}(k)=|t_\text{BB}|^2=
\frac{1}{2}+\frac{1}{2}\text{Re} \frac{a_+(k)a_-^{\ast}(k)}{a_+^{\ast}(k)a_-(k)}
\label{TBB-sym}
\end{equation} 
is equal to unity at resonances at $k=k_n$, where $k_n$ are determined from the equation
$\tan(2 k L)=-2k/\eta$.
For strong barriers $\eta\gg k, 1/L$, one finds 
\begin{equation}
k_n \simeq \frac{\pi n}{2L}\left(1-\frac{1}{\eta L}\right).
\label{kn-sym}
\end{equation}
The transmission coefficient is then described by a standard Breit-Wigner formula.

For an asymmetric setup with $\eta_R\neq \eta_L$, the transmission amplitude 
can still be written through $D(k)$ from Eq.~(\ref{DK}) as
\begin{equation}
t_\text{BB}(k)=2ik^2/D(k).
\label{tBB-asym}
\end{equation}
The transmission coefficient, 
\begin{eqnarray}
&&\left[\mathcal{T}_\text{BB}(k)\right]^{-1}\!
=\!1 +\!\frac{\eta_L^2 \eta_R^2}{8k^4} +\! \frac{\eta_L^2 + \eta_R^2}{4k^2} 
 \label{TBB-coeff-asym}
\\
&&+
 \frac{\eta_L \eta_R}{2k^2}\!\left[\!\left(1 - \frac{\eta_L \eta_R}{4 k^2}\right)\!\cos(4 k L)+  
    \frac{\eta_L + \eta_R}{2k} \sin(4 k L)
 \right],\notag
\end{eqnarray}
then also shows resonances at $k_n$ close to $\pi n/2L$ [in Eq.~(\ref{kn-sym}), $\eta$ is replaced
with $2\eta_R\eta_L/(\eta_R+\eta_L)$]. For strong barriers, $\mathcal{T}_\text{BB}$ is again described  by the Breit-Wigner formula, 
\begin{equation}
\mathcal{T}_\text{BB}(k\sim k_n)\simeq \frac{4\mathcal{T}_L\mathcal{T}_R}{64(k-k_n)^2L^2+(\mathcal{T}_L+\mathcal{T}_R)^2},
\label{transmission-BB-asym}
\end{equation}
where 
\begin{equation}
\mathcal{T}_{R,L}=\frac{4k^2}{4k^2+\eta^2_{R,L}}
\end{equation}
are the transmission coefficients of the individual barriers.
The height of the resonance is now always smaller than unity; for a strongly asymmetric setup,
\begin{equation}
\mathcal{T}_\text{max}\equiv\text{max}[\mathcal{T}_R,\mathcal{T}_L] \gg 
\mathcal{T}_\text{min}\equiv\text{min}[\mathcal{T}_R,\mathcal{T}_L],
\label{TmaxTmin}
\end{equation}
one gets
\begin{equation}
\mathcal{T}_\text{BB}(k=k_n)\simeq \frac{4\mathcal{T}_\text{min}}{\mathcal{T}_\text{max}}\ll 1.
\label{transmission-strong-asym}
\end{equation}
We will use the notation introduced in this section in Sec.~\ref{Sec:Double}, where a Fano state inside the double-barrier structure is considered.

\subsection{System with a Fano state and a tunnel barrier}
\label{sec:BF}
Another example of an effectively two-barrier structure is a quantum wire with a single tunnel barrier and a side-attached Fano state at some distance from it (see Ref. \cite{Kivshar} for similar engineered Fano systems). In such a structure, there is a peculiar effect of transmission enhancement by the Fano antiresonance, even for a very strong tunnel barrier. Indeed, one sees from Eq.~(\ref{refFano}) that in the energy band of width $\Gamma_0$ around the value of $E_k=\varepsilon_0$, the reflection from the Fano state is almost perfect, $r_F\simeq -1$. This implies that there is an energy at which the effective strength of the emerging Fano-barrier is exactly equal to the strength of an arbitrarily strong tunnel barrier. For such energy, we obtain a perfectly symmetric double-barrier structure. When this energy coincides with the level quantization energy in this double-barrier structure, the transmission coefficient for the whole system is unity. The energy dependence of the transmission amplitude $t_\text{BF}$ for the Barrier-Fano
structure is presented in the Appendix.

The perfect resonance in $\mathcal{T}_\text{BF}$ can be adjusted, e.g., by changing $\varepsilon_0$ with the external gate voltage, which is expected to be a much more precise way of fine tuning compared to the creation of two perfectly equal strong barriers. Notably, the ``transparency window'' of energies around the fine-tuned value emerges in a system consisting of two scatterers -- the tunnel barrier and the Fano level -- each of which individually \textit{suppresses} the transmission. A natural question is whether this phenomenon of transparency enhancement by the Fano antiresonance persists in more complex structures with low transmission, like an asymmetric double-barrier setup (\ref{TmaxTmin}), that cannot be directly mapped to the known systems characterized by a perfect transmission. As we will show in the remaining part of the paper, the answer to this question is positive.

\section{Double-barrier structure with a Fano state}
\label{Sec:Double}

\begin{figure}[t]
	\begin{center}
		\includegraphics[width=\linewidth]{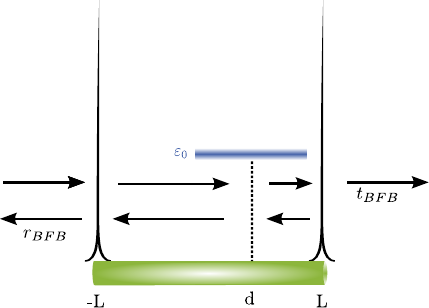} 
		\caption[Scattering off a Fano state in a system of two symmetric barriers]{Scattering off a Fano state in a system of two equal $\delta$-barriers located at $\pm L$. The coefficients of the wave functions are indicated. The strength of the barriers is quantified by $\eta$. The Fano state has the energy $\varepsilon_0$ and is located at position $d$. }
		\label{fig:TBMB}
	\end{center}
\end{figure}

In this section, we consider the structure with two tunnel barriers and the Fano state between them (Fig.~\ref{fig:TBMB}), where the localized level lies on top of a quasi-discrete spectrum of size quantization. It will be demonstrated that for strong barriers, a Fano state may lead to a strong transmission enhancement in a narrow energy window, similarly to the case of a single-barrier structure described in Sec.~\ref{sec:BF}. However, in contrast to that case, the effect of transparency enhancement for a two-barrier structure is far from being obvious, since the spectrum of the system is quasi-discrete from the outset, which can be expected to pose restrictions on the possibility of the fine-tuning of resonances. 

\subsection{Formalism}

If the 1D channel hosting the Fano state is tunnel-coupled to external leads, the resonant interference of electron waves scattered of the tunnel barriers becomes important.
With the Fano state side-attached to the region of the wire inside the double-barrier structure (Fig.~\ref{fig:TBMB}), 
the Hamiltonian of the system, 
\begin{align}
\nonumber
\hat{\mathcal{H}} & = \hat{\mathcal{H}}_0 + \varepsilon_0|\varphi\rangle \langle \varphi| \\&\nonumber
+\int_0^\infty dk\left(V_{-k} |\psi_{-k}\rangle \langle \varphi|+V_{-k}^{\ast} |\varphi\rangle \langle \psi_{-k}|\right) 
\\
&+ \int_0^\infty dk\left(V_{+k} |\psi_{+k}\rangle \langle \varphi|+V_{+k}^{\ast} |\varphi\rangle \langle \psi_{+k}|\right).
\label{hFano_barr}
\end{align}
is modified compared to Eq.~\eqref{hFano} by accounting for a quasi-discrete spectrum.
In Eq.~(\ref{hFano_barr}), $\hat{\mathcal{H}}_0$ is the Hamiltonian of the 1D wire with the two barriers. 
The corresponding eigenfunctions are given by
$$\psi_{\pm k}(x,\textbf{r}_\perp)=\phi_{\pm k}(x)\xi(\textbf{r}_\perp)$$ 
with $\phi_{\pm k}(x)$ given by Eqs.~\eqref{wf_r} and \eqref{wf_L}. The matrix elements $V_{\pm k}$ of the tunneling operator $\hat V$ are defined with respect to these eigenfunctions. The function $\xi(\textbf{r}_\perp)$ depends on the coordinate transverse to the wire. The energy $\varepsilon_0$ is the energy of the discrete level located at position $x=d$ with the corresponding wave function $\varphi(x,\textbf{r}_\perp)$. The functions $\psi_{\pm k}(x,\textbf{r}_\perp)$ and $\varphi(x,\textbf{r}_\perp)$ are orthogonal to each other because of a zero overlap in the transverse direction. 

Let us first analyze the case of symmetric setup with equal barriers characterized by $\eta$.
For a symmetric pointlike tunneling operator $\hat V$ and the Fano state located at position $x=d$, the coupling to left and right moving waves is given by
\begin{align}
V_{+k}&=V \left[a_+(k)\cos(kd)+a_-(k)\sin(kd)\right], 
\label{Vkp}
\\ V_{-k}&=V\left[a_+(k)\cos(kd)-a_-(k)\sin(kd)\right].
\label{Vkm}
\end{align}
Here, $a_+(k)$ and $a_-(k)$ are given by Eqs.~\eqref{a+} and \eqref{a-} and $V$ is a constant characterizing the strength of tunneling.

The equation for the eigenfunctions $\Phi_{+k}$ of the full Hamiltonian, corresponding to electrons moving through the system from left to right, can be written as
\begin{align}\label{trans_func}
\Phi_{+k}(x,\textbf{r}_\perp)&=\psi_{+k}(x,\textbf{r}_\perp)\\
&+\int G_{k}(x,\textbf{r}_\perp;x',\textbf{r}'_\perp){\hat V}\Phi_{+k}(x',\textbf{r}'_\perp) dx'd\textbf{r}'_\perp.
\nonumber
\end{align} 
Here, $G_{k}(x,\textbf{r}_\perp;x',\textbf{r}'_\perp)$ is the Green's function of the operator $H_0+\varepsilon_0|\varphi\rangle \langle \varphi|$.

To solve the self-consistent equation for the right-moving function Eq.~\eqref{trans_func}, we expand the full function of the system in terms of the eigenfunctions of the Hamiltonian without coupling. With the orthogonality of the functions $\psi_{-k,+k}(x,\textbf{r}_\perp)$ and $\varphi(x,\textbf{r}_\perp)$, we obtain (for brevity, we suppress the arguments $x,\textbf{r}_\perp$ in all eigenfuctions):
\begin{align}\label{trans_func2}
&\Phi_{+k}=\psi_{+k}+\frac{V_{-k}}{E_k-\varepsilon_0-\Sigma(k)}\varphi\\
&+\frac{V_{+k}^\ast}{E_k-\varepsilon_0-\Sigma(k)}\int_0^\infty dk'
\frac{V_{+k'}\psi_{+k'}+V_{-k'}\psi_{-k'}}{E_k-E_{k'}+i0}
\nonumber
\end{align}  
with  the self-energy
\begin{align}
&\Sigma(k)=\int_0^\infty dk'\frac{|V_{+k'}|^2+|V_{-k'}|^2}{E_k-E_{k'}+i0}.
\end{align}	
The real part of the self-energy, which is given by the principal-value integral, determines the position of the shifted resonance, $\varepsilon'_0(k)=\varepsilon_0+\delta\varepsilon_0(k)$, where
\begin{equation}
\delta\varepsilon_0(k)=\dashint_0^\infty dk'\frac{|V_{+k'}|^2+|V_{-k'}|^2}{E_k-E_{k'}+i0}.
\end{equation}
The imaginary part, as usual, describes the broadening of the Fano state,
\begin{equation}
\Gamma(k)=\frac{\pi m}{k}(|V_{+k}|^2+|V_{-k}|^2).
\label{Gammak-V}
\end{equation}
We are now prepared to calculate the transmission amplitude $t_\text{BFB}$ for the Barrier-Fano-Barrier (BFB) structure (Fig.~\ref{fig:TBMB}), which is done in the following section.

\subsection{Transmission coefficient}

We evaluate the integrals in Eq.~\eqref{trans_func2} at the pole of the denominator, since the other contributions decrease strongly with $x$. The amplitude of the outgoing wave for $x\gg L$ defines the transmission amplitude for the whole BFB structure:
\begin{align}
t_\text{BFB}(k)\!=\!t_\text{BB}(k)
\!-\!\frac{2i\pi m}{k}\,\frac{|V_{+k}|^2 t_\text{BB}(k)-V_{-k}V^{\ast}_{+k} r_\text{BB}(k)}
{E_k-\varepsilon'_0(k)+i \Gamma(k)}.  
\label{tBMB_func}
\end{align} 
Here $t_\text{BB}$ [coming from $\psi_{+k}$ and $\psi_{+k'}$ in Eq.~\eqref{trans_func2}] and $r_\text{BB}$ (coming from $\psi_{-k'}$) are the transmission and reflection coefficients of the double-barrier structure given by Eqs.~(\ref{TBB}) and (\ref{RBB}), respectively.

Using Eqs.~(\ref{Vkp}) and (\ref{Vkm}), we express the tunneling matrix elements in Eqs.~(\ref{Gammak-V}) and (\ref{tBMB_func}) explicitly through the functions $a_\pm(k)$ from Eqs.~(\ref{a+}) and (\ref{a-}). Next, combining the terms proportional to $t_\text{BB}$, we use the identity that follows from Eqs.~(\ref{TBB}) and (\ref{RBB}),
\begin{align}
t_\text{BB}(k)\, \text{Re}\left\{a_+(k)a_-^\ast(k)\right\}
= i\, r_\text{BB}(k)\, \text{Im}\left\{ a_+(k)a_-^\ast(k)\right\},
\end{align}
to re-write Eq.~\eqref{tBMB_func}, leading to:
\begin{widetext}
\begin{align}
&t_\text{BFB}(k)=\frac{[E_k-\varepsilon_0-\delta\varepsilon_0(k)]t_\text{BB}(k)-i\Gamma_0(k)
[|a_+(k)|^2\cos^2(kd)-|a_-(k)|^2\sin^2(kd)]r_\text{BB} }
{E_k-\varepsilon_0-\delta\varepsilon_0(k)+i\Gamma_0(k)[|a_+(k)|^2\cos^2(kd)+|a_-(k)|^2\sin^2(kd)]},
\label{tBFB-intermed}
\end{align}
where we have introduced $\Gamma_0(k)=2\pi m |V|^2/k$. 
Expressing the shifted resonance energy as 
\begin{align}
\delta\varepsilon_0(k)&=
|V|^2\!\dashint_0^\infty\!\!dk'\frac{|a_+(k')|^2\cos^2(k'd)+|a_-(k')|^2\sin^2(k'd)}{E_k-E_{k'}},
\label{realpart_Fano}
\end{align} 
\end{widetext}
we use the following exact relations for the real and imaginary part 
of the fractions of $a_+(k)$ and $a_-(k)$:
\begin{align}
\label{ReIma+}
\text{Re}\frac{a_+(k)}{a_-(k)}&=-i|a_+(k)|^2\frac{r_\text{BB}}{t_\text{BB}},
\quad \text{Im}\frac{a_+(k)}{a_-(k)}=-|a_+(k)|^2,
\\
\quad
\text{Re}\frac{a_-(k)}{a_+(k)}&=-i|a_-(k)|^2\frac{r_\text{BB}}{t_\text{BB}},
\quad 
\text{Im}\frac{a_-(k)}{a_+(k)}=|a_-(k)|^2.
\label{ReIma}
\end{align}
We can now directly evaluate the principal value integral~\eqref{realpart_Fano} via the Kramers-Kronig relation,
\begin{align}
\text{Re}\,F(x)\ {\overset{\text{K.K.}}{=}}\ 
\frac{2}{\pi}\dashint_0^\infty\frac{t\,\text{Im}\,F(t)}{t^2-x^2}dt.
\end{align}
Applying this to Eq.~\eqref{realpart_Fano} results in 
\begin{align}
&\dashint_{0}^{\infty} dk'\frac{|a_+(k')|^2}{E_k-E_k'}
=-\dashint_{0}^{\infty} dk'\frac{k'}{E_k-E_k'}\text{Im}\,\frac{a_+(k')}{k'a_-(k')}\nonumber\\
&\ {\overset{\text{K.K.}}{=}}-\pi m\text{Re}\,\frac{a_+(k)}{k a_-(k)}=i\frac{\pi m}{k}|a_+(k)|^2\frac{r_\text{BB}}{t_\text{BB}},
\end{align}
and, similarly,
\begin{align}
\dashint_{0}^{\infty} dk'\frac{|a_-(k')|^2}{E_k-E_k'}
=-i\frac{\pi m}{k}|a_-(k)|^2\frac{r_\text{BB}}{t_\text{BB}}.
\end{align}

The transmission amplitude can then be expressed in a very compact form, reading
\begin{align}
\label{tBMBCompact}
t_\text{BFB}(k)=\frac{E_k-\varepsilon_0}
{E_k-\varepsilon_0-\Sigma(k)}t_\text{BB}(k)
\end{align}
with
\begin{eqnarray}\label{sigma_symm}
\Sigma(k)=\Gamma_0\left[\frac{a_+(k)}{a_-(k)}\cos^2(kd)-\frac{a_-(k)}{a_+(k)}\sin^2(kd)\right].
\end{eqnarray}
The transmission coefficient $\mathcal{T}_\text{BFB}(k)=|t_\text{BFB}(k)|^2$ thus reads:
\begin{equation}
\mathcal{T}_\text{BFB}(k)
 =\frac{ (E_k-\varepsilon_0)^2 \mathcal{T}_\text{BB}(k)}{\left[E_k-\varepsilon_0-\text{Re}\Sigma(k)\right]^2
	+\left[\text{Im}\Sigma(k)\right]^2}.
	\label{TBFBK}
\end{equation}

Note that, somewhat surprisingly, the transmission coefficient $\mathcal{T}_\text{BFB}$ 
is exactly zero at the original energy 
$\varepsilon_0$ of the Fano state, despite the fact that the matrix elements of the tunneling 
coupling between the localized state and the wire are now momentum dependent. In a continuum spectrum, such
a momentum dependence leads to the shift of the energy where the transmission coefficient vanishes,
see Eqs.~(\ref{trFano}) and (\ref{dE}). At the same time, the resonance energy in the denominator is shifted
from the value $\varepsilon_0$, similarly to Eqs.~(\ref{trFano}) and (\ref{dE}) for the momentum dependent coupling. This asymmetry between the numerator and denominator in Eq.~(\ref{TBFBK}), introduced by the combination of the Fano destructive interference and the level quantization, is a distinct feature of the double-barrier Fano setup.

\begin{figure}
  \centering
\includegraphics[width=\linewidth]{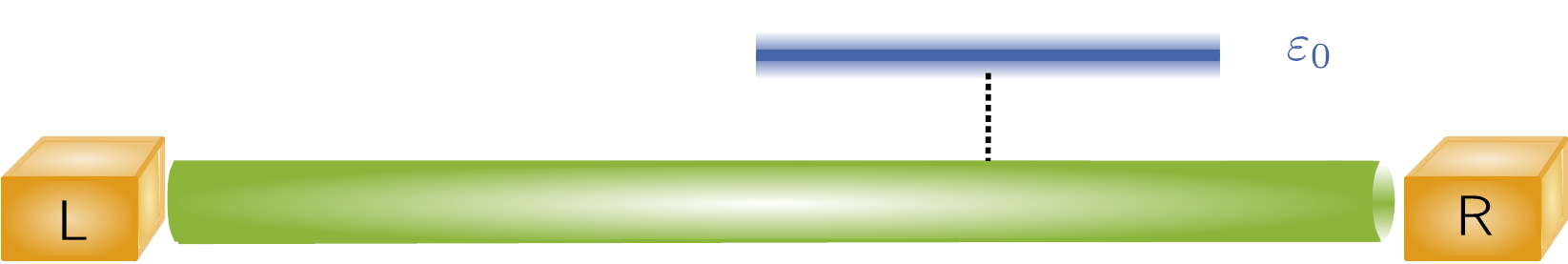} \\[0.2cm]
  \includegraphics[width=\linewidth]{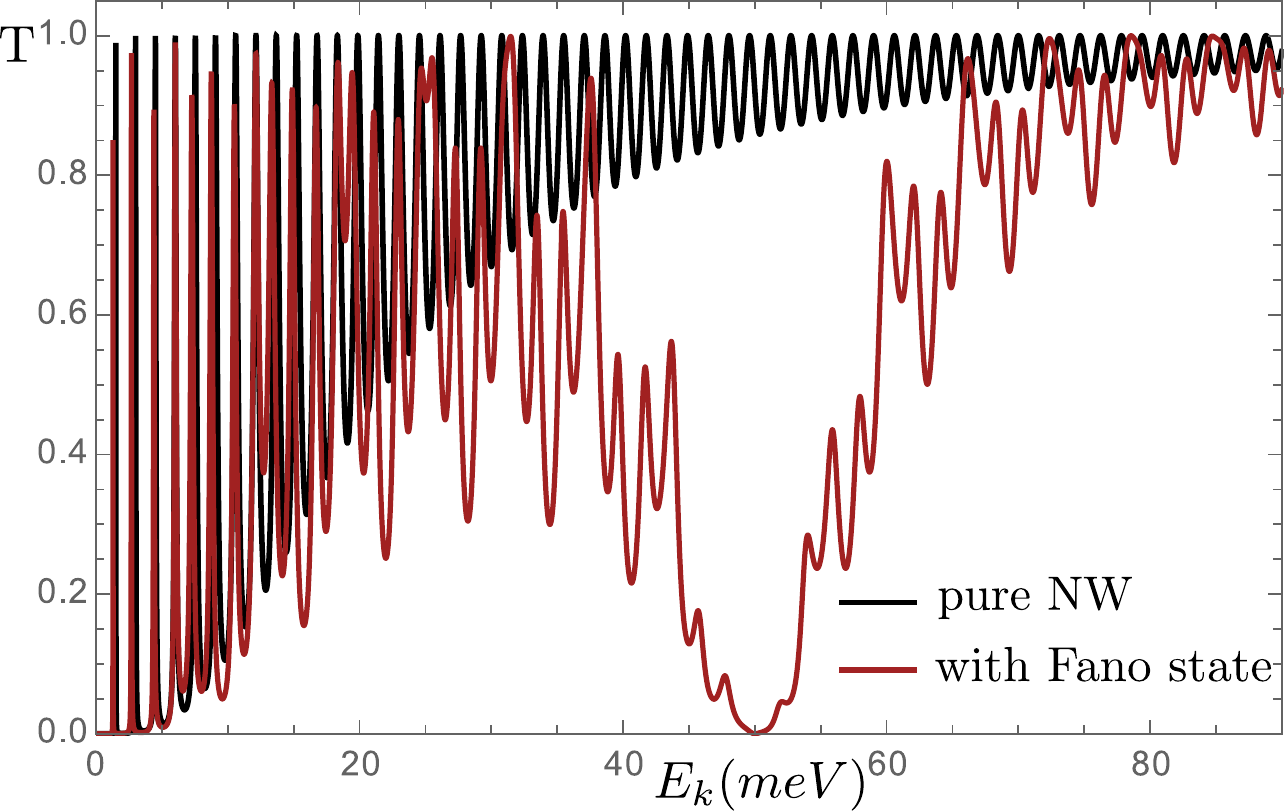}
   \caption[Transmission coefficient in a setup of an 1D channel tunnel-coupled to the leads in presence of a Fano state]{Upper panel: Schematic illustration of the setup with an 1D channel coupled to the leads by tunneling barriers in the presence of a Fano state. Lower panel: Transmission coefficient for this setup as a function of the carrier energy $E_k$ for a nanowire (NW) with symmetric barriers and strong coupling to the Fano state (located at $d/L=0.6$). The red curve shows the transmission $\mathcal{T}_\text{BFB}$ for the structure with barriers and a Fano state, the black one is for two barriers without the Fano state 
   ($\mathcal{T}_\text{BB}$). The chosen parameters roughly correspond to those in the experimental setup of Ref. \cite{Krainov}. The energy of the localized state is $\varepsilon_0 = 50$~meV; the transmission coefficients across the contacts to the leads are characterized by the broadening of size-quantized levels at the Fermi energy, $\eta_\text{R}=\eta_\text{L} = 3.16$~meV; the hybridization of conducting electrons with the Fano state is characterized by $\Gamma_0 = 30$~meV. }
  \label{fig:TBMB_quant}
\end{figure}

The result (\ref{TBFBK}) is illustrated in Figs.~\ref{fig:TBMB_quant} and \ref{fig:resonance}. 
For a strong coupling of the Fano state to the wire, when the Fano-state broadening exceeds the level spacing in the double-barrier setup, $\Gamma_0\gg \Delta=E_n-E_{n-1}$ (where $E_n,E_{n-1}$ are the levels around $\varepsilon_0$), 
the transmission is suppressed in the region of energies 
$(\varepsilon_0 - \Gamma_0, \varepsilon_0 + \Gamma_0)$, similarly to the case without barriers, Sec. \ref{Sec:Fano}. For sufficiently weak barriers, $\mathcal{T}_{R,L}\sim 1$, the only difference compared with the continuum case is in the modulation of the transmission coefficient by the double-barrier resonances. These resonances for equal barriers 
are now not perfect, in contrast to the case without the Fano state, Sec.~\ref{Sec:Barriers}.

Remarkably, for strong barriers, $\mathcal{T}_{R,L}\ll 1$, narrow resonances appear through an interplay of the Fano interference and the size quantization of the spectrum in the double-barrier structure, see Fig.~\ref{fig:resonance}. In the following section, we will analyze this phenomenon in detail.

\section{Emergent resonance}
\label{Sec:Resonance}

\subsection{Equal barriers}

\begin{figure}
	\begin{center}
		\includegraphics[width=\linewidth]{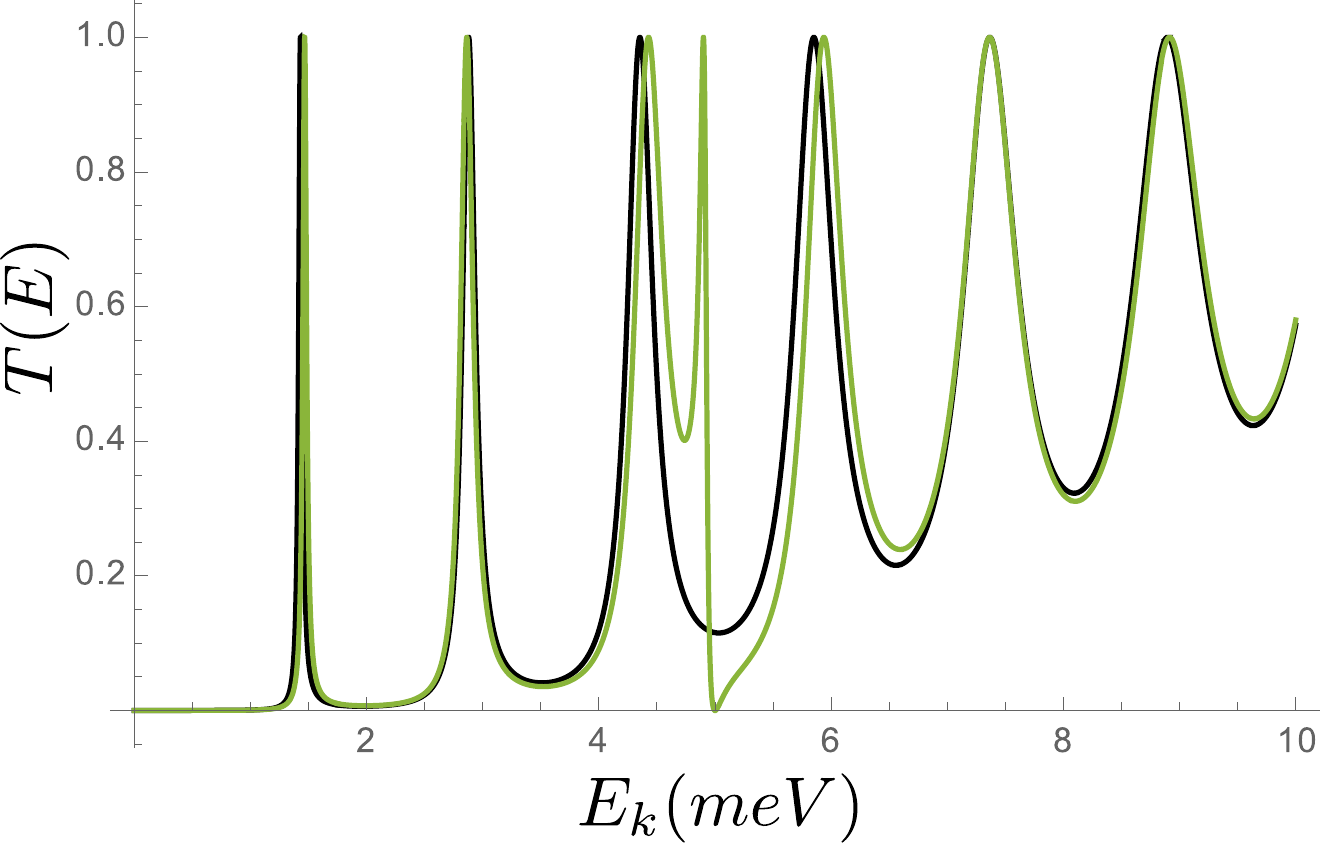} 
		\caption{Transmission coefficient of the double-barrier setup as a function of the carrier energy $E_k$ for symmetric barriers and the energy of Fano state located in the region of low energies, where the barriers are strong. The green curve shows the transmission $\mathcal{T}_\text{BFB}$ for the structure with barriers and the Fano state, the black one shows the transmission for two barriers without the 
side-attached state ($\mathcal{T}_\text{BB}$). 
The hybridization of conducting electrons with the Fano state located at $d/L=0.2$ is characterized by $\Gamma_0 = 1$~meV; the energy of the localized state is $\varepsilon_0 = 5$~meV; the transmission coefficients across the contacts to the leads are the same as in Fig. \ref{fig:TBMB_quant}.
}
		\label{fig:resonance}		
	\end{center}
\end{figure}

We start with the case of equal barriers considered in the previous section. We assume that $\mathcal{T}_R=\mathcal{T}_L\ll 1$, otherwise the transmission coefficient $\mathcal{T}_\text{BFB}$ is similar to that without barriers (continuum spectrum). In Fig.~\ref{fig:resonance}, we show the transmission coefficient of the Barrier-Fano-Barrier structure with the same parameters of the barriers as in Figs.~\ref{fig:TBMB_quant}, but with the energy of Fano state located in the region of low energies where barriers are sufficiently  strong ($\varepsilon_0=5$meV instead of 50meV).
As we see from Fig.~\ref{fig:resonance}, in the wire with quasi-discrete levels, a narrow resonance emerges in the vicinity of the energy $\varepsilon_0$ of the localized state where the transmission is strictly zero. Below we will determine the conditions for the emergence of a resonance and analyze its properties.

It is instructive to compare the structure of the obtained transmission coefficient $\mathcal{T}_\text{BFB}$, Eq.~(\ref{TBFBK}), with that of the conventional Fano resonance, Eq.~(\ref{Fano-q}).
Both expressions describe an asymmetric lineshape, with both a zero and a maximum present. 
However, as already emphasized above, $\mathcal{T}_\text{BFB}$ is exactly zero at the original position of the resonant level $\varepsilon_0$. At the same time, at any finite $q$ in Eq.~(\ref{Fano-q}), the zero is shifted from $\varepsilon_0$ in the conventional case. Thus, the combined effect of the Fano interference and the size quantization of the spectrum substantially modifies the conventional result (\ref{Fano-q}).

From Eq.~(\ref{TBFBK}), we see that the position of the resonance is given by the condition
\begin{align}
E_k=\varepsilon_0+\text{Re}\Sigma(k),
\end{align}
as in the conventional Breit-Wigner formula,
and the width of the resonance is determined by $\text{Im}\Sigma(k)$. 
Thus, both the position and the width of the resonance depend on the momentum. 
For small $\Gamma_0\ll \Delta$, the value of the momentum in $\Sigma(k)$ can be approximated by the momentum corresponding to the of the Fano state $k_0$.  
This further implies that the height of the resonance is given by
	\begin{align}
&\mathcal{T}_{\text{BFB},\text{res}}=\mathcal{T}_\text{BB}\left.\frac{\left[\text{Re}\Sigma(k)\right]^2}{\left[\text{Im}\Sigma(k)\right]^2}\right|_{k=k_0}.
\label{height_Res1}
\end{align}
For equal barriers, this formula takes the form
\begin{align}
&\mathcal{T}_{\text{BFB},\text{res}}=\mathcal{R}_\text{BB}\\
&\times
\left.\left[\frac{|a_+(k)|^2\cos^2(kd)-|a_-(k)|^2\sin^2(kd)}{|a_+(k)|^2\cos^2(kd)+|a_-(k)|^2\sin^2(kd)}\right]^2\right|_{k=k_0}.
\notag
	\end{align}	
This shows that in the fully symmetric setup, when the Fano state is coupled to the middle of the segment between the barriers, $d=0$, the height of the peak is given by the reflection coefficient of the double-barrier system:
\begin{align}
\mathcal{T}_{\text{BFB},\text{res}}
=\mathcal{R}_\text{BB}, \qquad d=0.
\end{align}
It is clearly seen that for strong barriers, $\eta \gg k$, the transmission coefficient can be very close to unity in the energy window where the transmission through the double-barrier system is suppressed by the destructive interference of waves reflected from the barriers. 

The height of the resonant peak for an arbitrary position of the side-attached state can be approximated for strong barriers, $\eta \gg k$ (and keeping $|a_+|, |a_-| \ll 1$) as
\begin{align}
\mathcal{T}_{\text{BFB},\text{res}}
=\mathcal{R}_\text{BB}\left.\left[\frac{\cos(2kL)-\cos(2kd)}{\cos(2kL)\cos(2kd)-1}\right]^2\right|_{k=k_0},
\label{height_Res}
\end{align}
defining the momenta where only the antiresonances can appear. 
Thus, for equal strong barriers, the transmission coefficient at the resonance can reach unity (perfect transmission). 
The corresponding width of the resonance is given by
\begin{align}
\Gamma_\text{res}&=-\text{Im}\Sigma(k_0)\notag
\\
&=\Gamma_0\frac{k^2}{\eta^2}\left[\frac{\cos^2(k_0d)}{\cos^2(k_0L)}+\frac{\sin^2(k_0d)}{\sin^2(k_0L)}\right]
\sim \Gamma_0 \sqrt{\mathcal{T}_\text{BB}},
\label{Gres-equal}
\end{align}
and can be much smaller than the width of the Breit-Wigner peaks in the transmission coefficient 
$\mathcal{T}_\text{BB}$. 


\begin{figure}[ht]
	\begin{center}
		\includegraphics[width=0.95\linewidth]{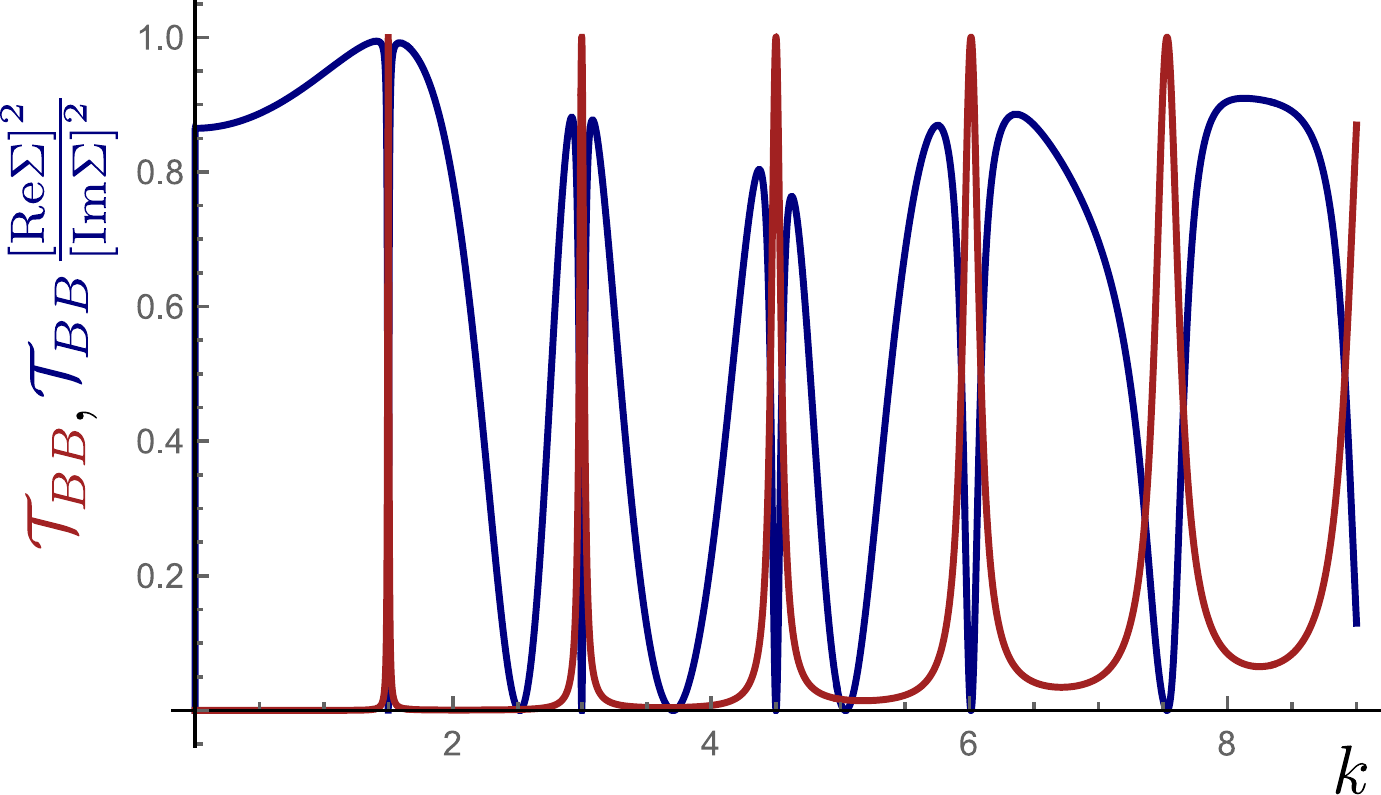} 
		\caption{Transmission coefficient for a symmetric double-barrier structure (red curve) and the height of the Breit-Wigner resonances, Eq.~\eqref{height_Res}, for the position of the Fano state at $d/L=0.2$ (dark blue curve) and the barrier strength $\eta L=40\pi$. }
		\label{fig:resonance_pos_arb}
	\end{center}
\end{figure}


\subsection{General case}

Importantly, a narrow resonant peak with an almost perfect transmission also emerges for an asymmetric double-barrier setup. In this case, the transmission coefficient $\mathcal{T}_\text{BB}$ for the system without a Fano state is governed by the asymmetry of the structure and thus can be arbitrarily small for all energies, see Eqs.~(\ref{transmission-BB-asym}) and (\ref{transmission-strong-asym}). However, even in this case, the presence of the Fano state can lead to a strong resonance with the transmission coefficient that can be again of the order of unity.

The expression for the transmission amplitude 
is given by Eq.~(\ref{tBMBCompact}),
with $t_\text{BB}(k)$ from Eq.~(\ref{tBB-asym}) and
$\Sigma(k)$ defined as
	\begin{align}
	&\Sigma_\text{asym}(k)=\frac{2\Gamma_0}{a_{l+}a_{r-}+a_{l-}a_{r+}}
	\Big[a_{r+}a_{l+}\cos^2(kd) \label{Sigma-asym}
		\\ 
	&-a_{r-}a_{l-}\sin^2(kd)
	+(a_{l+}a_{r-}\!-a_{l-}a_{r+})\cos(kd)\sin(kd)\Big].
	\notag
	\end{align}
Here, $a_{r\pm}$ and $a_{l\pm}$ are defined in Eqs.~(\ref{aR+}) and (\ref{aR-}). 
In the limit of the equal barriers, the result for the self-energy for the equal barriers, Eq.~\eqref{sigma_symm}, is recovered.
The result for the transmission amplitude $t_\text{BFB}$ can be cast in a compact form 
through the reflection and transmission amplitudes of individual barriers (here the transmission from left to right is assumed):
\begin{gather}
t_\text{BFB} = \frac{(E_k - \varepsilon_0)\ t_\text{BB}}
{E_k - \varepsilon_0 + i \Gamma_0 
\frac{[1+r_L \e^{2 i k (L+d)}][1 + r_R \e^{2 i k (L - d)}]}{1- r_L r_R \e^{4 i k L}}}.
\label{tLFR-result}
\end{gather}
For an alternative derivation (based on the scattering-matrix approach) of the general formula for $\mathcal{T}_{\text{BFB}}$, see Ref.~\cite{footnote}. It follows directly from Eq.~\eqref{tLFR-result} that the emergence of the Fano-induced resonance requires that at least one of the barriers is strong, $\mathcal{T}_\text{min}\ll 1$. When the second barrier is weak, the situation is similar to that described in Sec.~\ref{sec:BF} (see also Appendix). When both barriers are strong, the spectrum of the structure is quasi-discrete. In what follows, we analyze the transmission through the whole structure under this condition.


\begin{figure}[h]
	\begin{center}
		\includegraphics[width=0.95\linewidth]{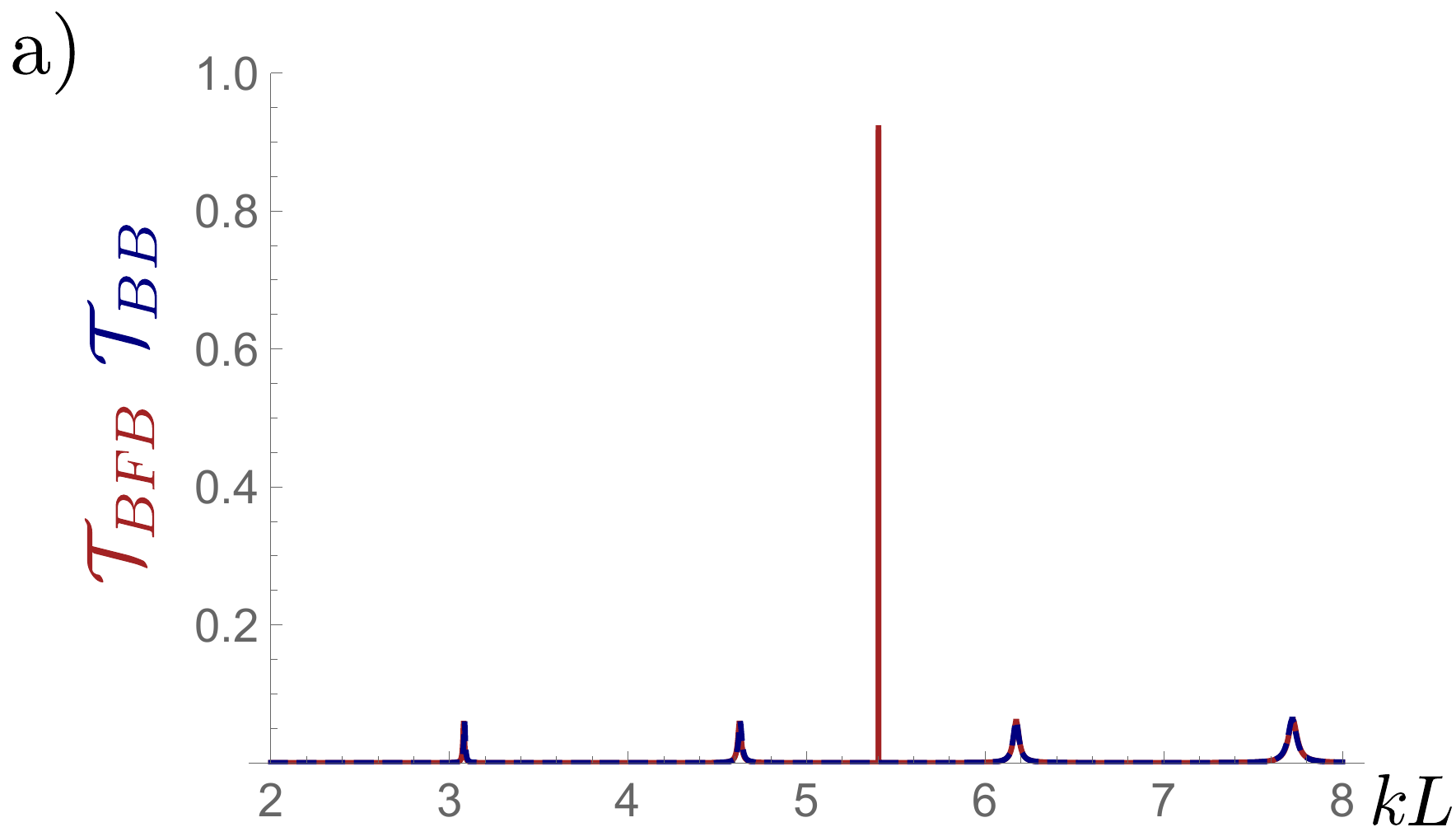} \\
				\includegraphics[width=0.95\linewidth]{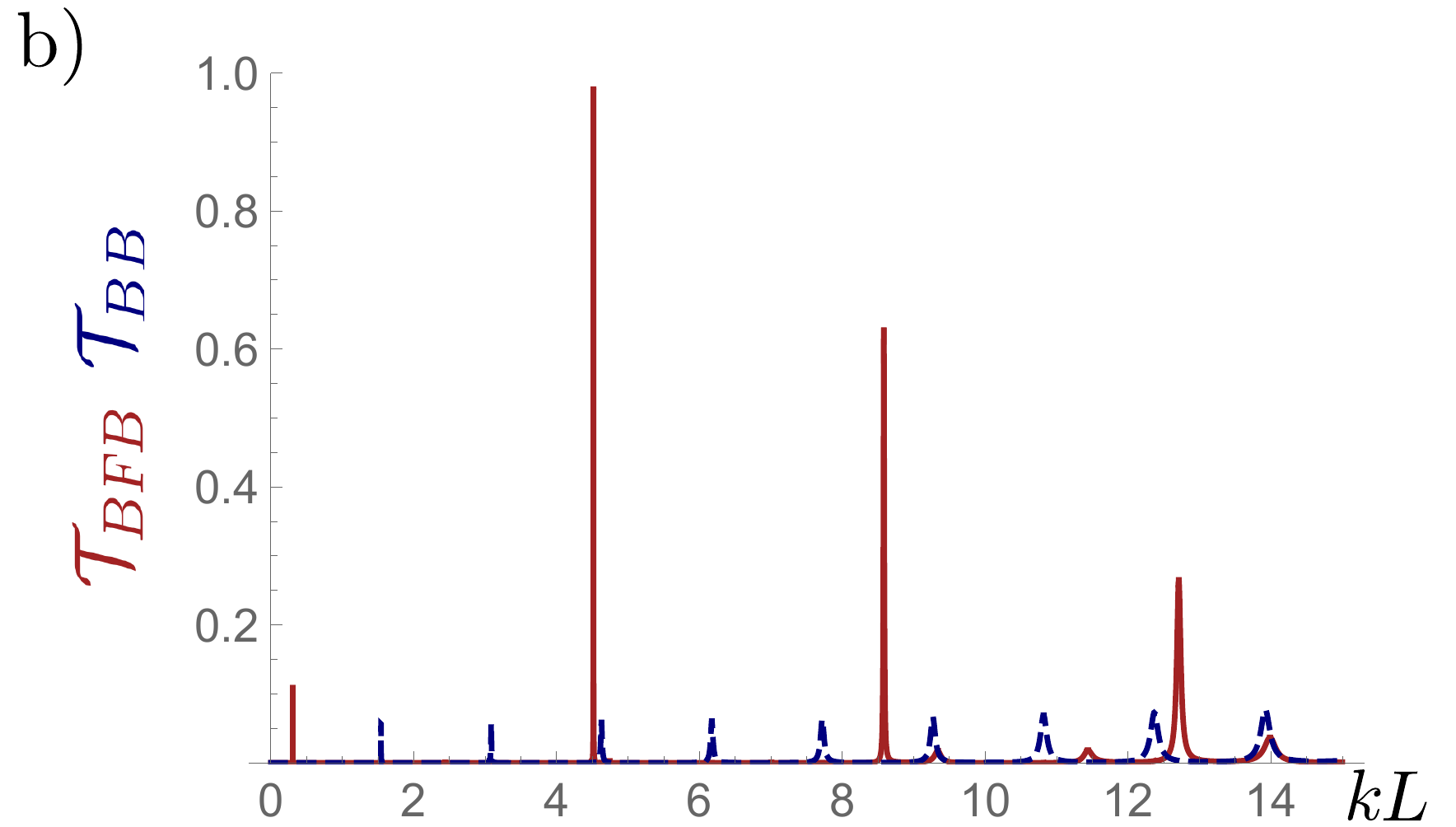}
\caption{Transmission coefficient for asymmetric double-barrier structures without (blue dashed line) and with (red line) a Fano state for weak [$\Gamma_0\ll \Delta$, panel (a)] and strong [$\Gamma_0\gg \Delta$, panel (b)] coupling between the wire and the Fano state. 	
In both panels, $k_0 L = 5.4$, $\eta_L L = 30$ and $\eta_R L = 250$.
Panel (a): $\Gamma_0 m L^2 / \hbar^2 = 0.1$ and $d=0.45 L$.
Panel (b): $\Gamma_0 m L^2 / \hbar^2 = 100$ and $d=0.3 L$.}
		\label{fig:resonance_strong_arb}
	\end{center}
\end{figure}


The emergent resonances for an asymmetric strong-barrier setup are shown in Fig.~\ref{fig:resonance_strong_arb} for weak and strong coupling between the wire and the Fano state.
We observe that strong narrow resonances emerge for an arbitrary tunneling coupling between the Fano state.
For weak coupling (Fig.~\ref{fig:resonance_strong_arb}a), the emergent resonance is located close to the Fano-state energy $\varepsilon_0$, while away from $\varepsilon_0$ the transmission through the structure is almost not affected by the Fano state. In the opposite limit (Fig.~\ref{fig:resonance_strong_arb}b), several new peaks emerge within the energy window $(\varepsilon_0-\Gamma_0,\varepsilon_0+\Gamma_0)$, while the peaks in $\mathcal{T}_\text{BB}$ in this window are suppressed by the presence of the side-attached state.
Indeed, as we demonstrate in Appendix, the emergent resonances correspond to energies $E_k$ obeying 
the relation $E_k\simeq \varepsilon_0-\Gamma_0 \cos(2kd)$. For $\Gamma_0\ll \Delta$, there is only one solution for $E_k$ that satisfies Eq.~(\ref{emergent-spectrum}), whereas for $\Gamma_0\gg \Delta$ there are many emergent resonances.

The analysis presented in Appendix yields
Eq.~(\ref{TBFB-strong-asym}) away from the quantization levels of the double-barrier system without the Fano state. This formula shows that the resonance width (defined as the width of the peak at the half of the height) is given by
\begin{equation}
\Gamma_\text{res}\sim \text{min}[\Gamma_0,\Delta] \mathcal{T}_\text{min}.
\label{Gres-asym}
\end{equation}
Here $\mathcal{T}_\text{min}$ and $\mathcal{T}_\text{max}$ are the transmission coefficients of the individual barriers [see Eq.~(\ref{TmaxTmin})].
For equal barriers, Eq.~(\ref{Gres-asym}) can be written through $\mathcal{T}_\text{BB}$, as in Eq.~(\ref{Gres-equal}).

Using 
$\mathcal{T}_\text{BB} \approx \mathcal{T}_L \mathcal{T}_R / [4 \sin^2(2kL)]$
away from the resonances in $\mathcal{T}_\text{BB}$
and Eq.~(\ref{TBFB-res-asym}) from Appendix, we express the transmission coefficient at the emergent resonance as follows:
\begin{align}
\mathcal{T}_\text{BFB,res}\!&\approx\! \mathcal{T}_L \mathcal{T}_R 
\label{TBFBres-general}
\\
&\times \!
\left\{\!\frac{2\sin[k_\text{res}(L\!-\!d)] \sin[k_\text{res}(L\!+\!d)]}
{\mathcal{T}_L  \sin^2[k_\text{res}(L\!-\!d)] + \mathcal{T}_R  \sin^2[k_\text{res}(L\!+\!d)]}\!\right\}^2.
	\notag
\end{align}
Here $k_\text{res}$ is found from the minimum of the denominator of Eq.~(\ref{TBFB-strong-asym}).

\begin{figure}[ht]
	\begin{center}
		\includegraphics[width=\linewidth]{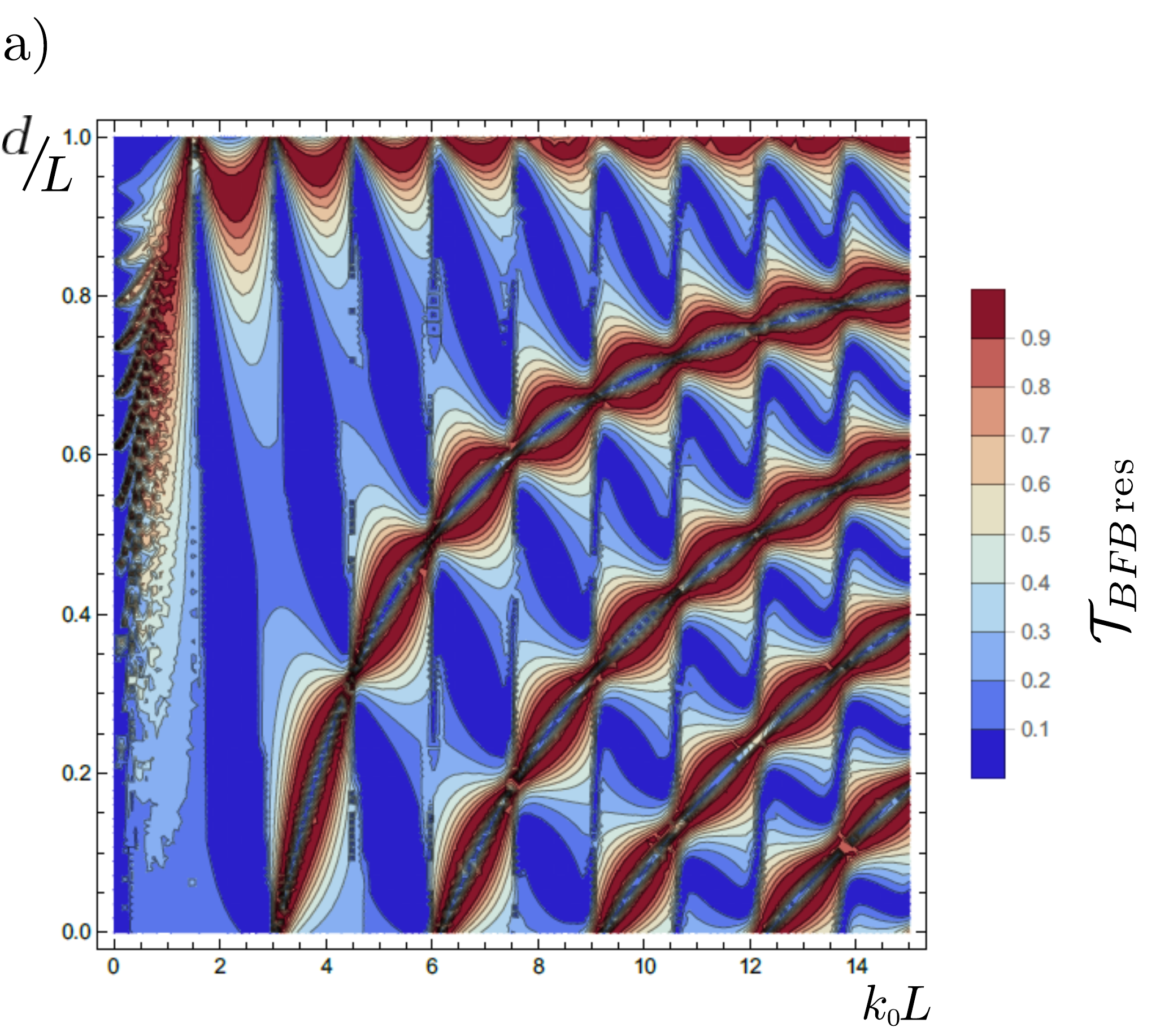} 
		\includegraphics[width=\linewidth]{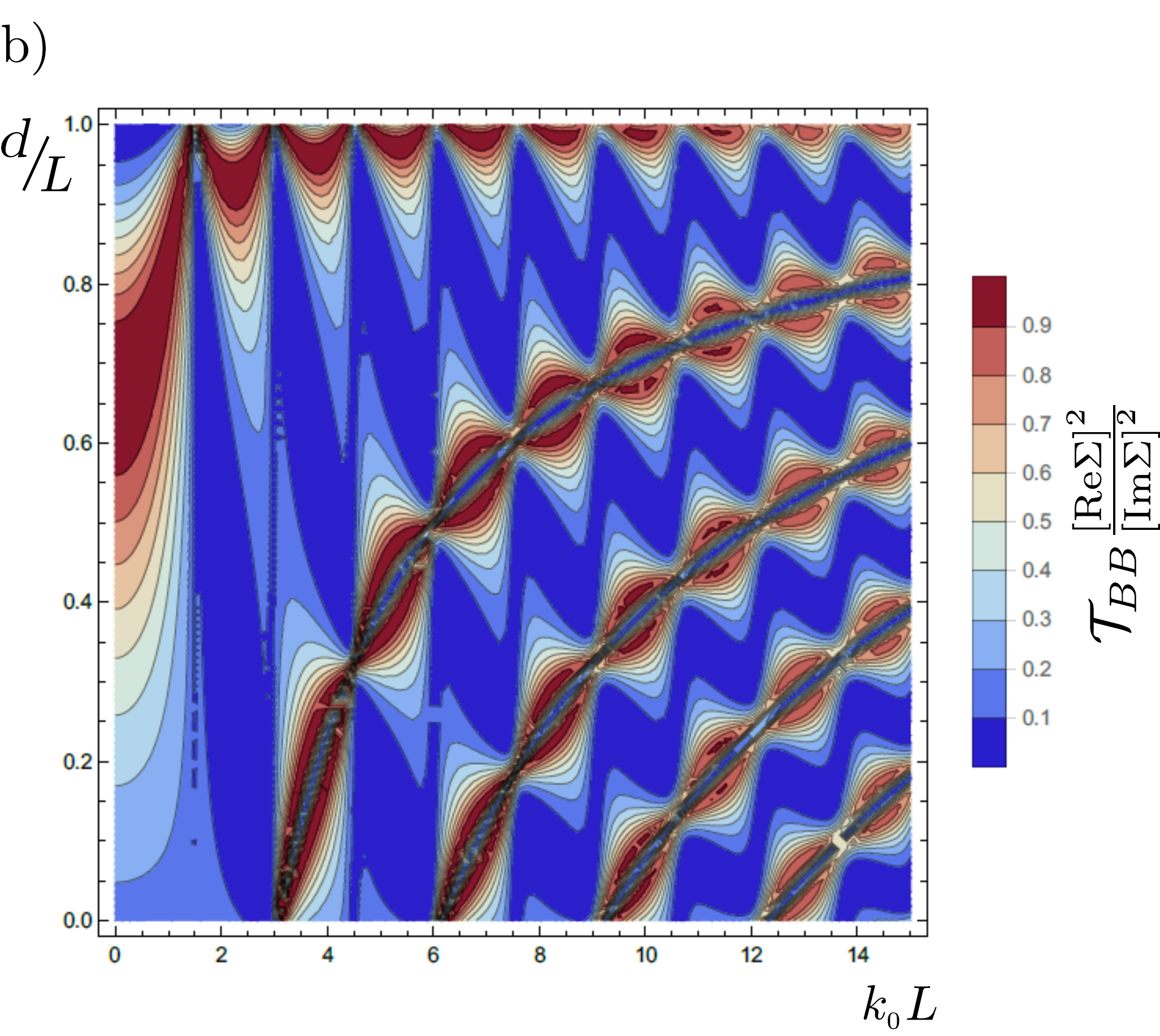}
		\caption{
		 The contour plot displaying the height of the resonance in terms of the position of the Fano state $d/L$ and the momentum of the Fano state $k_0$ evaluated numerically (a) with the exact formula  and (b) with the approximation \eqref{height_Res1}. The barriers are characterized by $\eta_L L=10$, 
$\eta_R L = 50$ and the coupling strength for tunneling between the Fano state and the wire is $\Gamma_0 = \hbar^2/m L^2$. 
		}
		\label{contour_height}
	\end{center}
\end{figure}

Let us now focus on the case of weak coupling ($\Gamma_0\ll \Delta$), where, according to Eq.~(\ref{emergent-spectrum}), one can set $k_\text{res}\to k_0$.
A resonance with the height of order of unity can be then found around the lines
\begin{equation}
k_0 (L \pm d) \simeq \pi n.
\label{res-condition}
\end{equation} 
Note that exactly on those lines the height of the 
emergent resonance vanishes, but 
quickly becomes of the order of unity in the close vicinity of the lines (the resonant region is determined
by $\mathcal{T}_{R,L}$). Specifically, the position of the Fano center should be adjusted with the accuracy 
\begin{equation}
\delta d \sim (1/k_0) \sqrt{\mathcal{T}_\text{min}/\mathcal{T}_\text{max}}
\label{res-accuracy}
\end{equation}
to obtain the height of the order of unity.

\begin{figure}
	\begin{center}
		\includegraphics[width=0.95\linewidth]{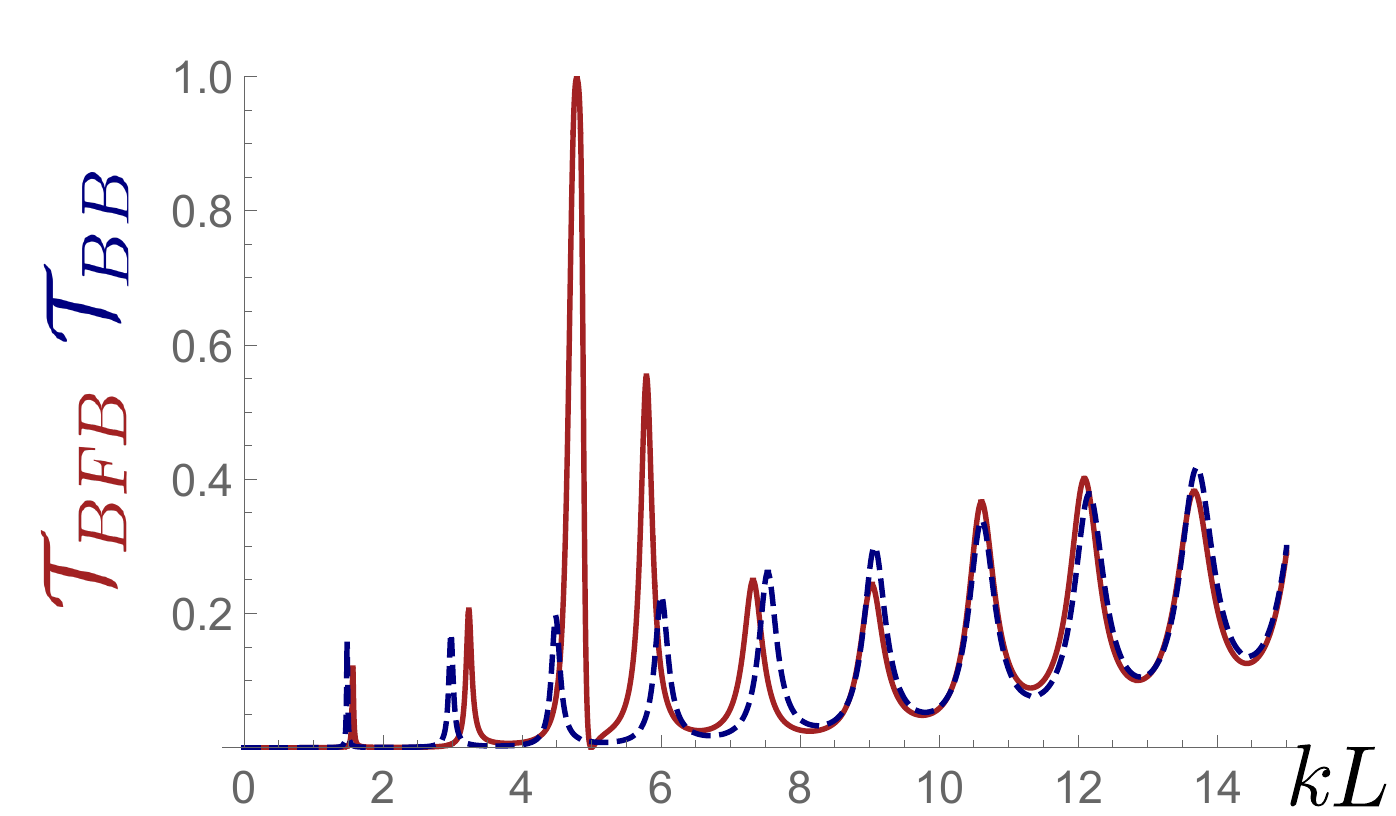} \\
		\caption{
Transmission coefficient for the asymmetric double-barrier structure at intermediate coupling 
($\Gamma_0 \sim \Delta$) without (blue dashed curve) and with (red curve) a Fano
state located at $d = 0.4 L$ with the level at $k_0 L = 5$. The barriers are characterized by 
$\eta_L L = 10$, $\eta_R L = 50$ and the coupling strength for tunneling
between the Fano state and the wire is $\Gamma_0 = \hbar^2/2 m L^2$ (same barriers and coupling as in Fig. \ref{contour_height}).}
		\label{fig:resonance_strong_arb_3}
	\end{center}
\end{figure}

Both the height and width of the emergent resonance oscillate with varying $k$, $k_0$, $d$, and $L$.
This is demonstrated in Fig.~\ref{contour_height} where the height of the resonance is displayed in a contour plot in terms of the position of the Fano state $d$ and the momentum of the Fano state $k_0$. Specifically, Fig.~\ref{contour_height}(a) shows the numerically obtained height of the resonance and Fig.~\ref{contour_height}(b) illustrates the validity of approximation \eqref{height_Res1}. 
In these plots, we clearly see the zeros of $\mathcal{T}_{\text{BFB}}$ along the positions of resonances in $\mathcal{T}_{\text{BB}}$, as well as along the curves defined by Eq.~\eqref{res-condition}, where the real part of the self-energy vanishes. The regions of maxima in $\mathcal{T}_{\text{BFB}}$ (dark red regions),
where the strongest transparency enhancement occurs, are located in the vicinity of these curves. 
An example of the $\mathcal{T}_{\text{BFB}}$
profile at the point in this plane corresponding to a strong emergent resonance ($d = 0.49 L$, $k_0 L = 5.6$) is shown in Fig. \ref{fig:resonance_strong_arb_3}.

For the parameters chosen in Fig.~\ref{contour_height} and the presented range of $k$, the maximum values of the transmission coefficient without the Fano state, $\mathcal{T}_{\text{BB}}$, range from $0.16$ to $0.4$. The height of the emergent resonance in $\mathcal{T}_{\text{BFB}}$ exceeds the peaks in $\mathcal{T}_{\text{BB}}$ in approximately a half of the area of the shown parameter plane.  In general, the region where the height of emergent resonance is not smaller than the height of resonances in $\mathcal{T}_{\text{BB}}$ is always comparable to the total area in the parameter plane.
Indeed, for a fixed value of $k_0L$, Eq.~\eqref{TBFBres-general} yields a magnitude of the resonance of the order of $\mathcal{T}_{\text{BB}}$ multiplied by a function of $d$ constructed out of trigonometric functions.


The probability of finding a strong resonance of order unity (the area of red areas in Fig.~\ref{contour_height}) can be estimated as follows. There is a periodic dependence on $k_0L$ in Eq.~\eqref{TBFBres-general}. The width of each red area is $\delta_{k_0L} \sim 1$ and the height is $\delta_{d/L} \sim (1/k_0L) \sqrt{\mathcal{T}_\text{min}/\mathcal{T}_\text{max}}$, see Eq.~\eqref{res-accuracy}. 
Thus, the probability of having a strong 
resonance is given by $\sqrt{\mathcal{T}_\text{min}/\mathcal{T}_\text{max}}$. With increasing asymmetry between the barriers, this area shrinks, but the height of the resonances in $\mathcal{T}_{\text{BB}}$ also decreases. As a result, the height of the emergent resonances in $\mathcal{T}_{\text{BFB}}$ typically competes with the height of resonances in $\mathcal{T}_{\text{BB}}$, and exceeds the latter in roughly half of the parameter range, as discussed above.

Given that the transmission at all other energies is suppressed by the asymmetry of the setup, the resonances shown in Fig.~\ref{fig:resonance_strong_arb} can be employed for 
the narrow-band filtering of the electronic waves. Since the position of the resonance is determined by the Fano-state energy, such a transport filter is tunable by electrostatic gates. In contrast to the tunneling through the localized state, almost perfect transmission through the structure does not require fine-tuning the symmetry of the setup. The height of the transmission peak can be additionally controlled by changing the location of the side-attached state.

\section{Conclusion}

In this paper, we have shown that the conventional picture of resonant tunneling across a double-barrier structure is strongly modified by a Fano state side-attached between the barriers. When at least one of the barriers is strong, the presence of the Fano state may lead to the emergence of narrow strong resonances. The width and height of emergent resonances depend on the energy and position of the Fano state, the strength of the coupling between the wire and the Fano center, the strength of the barriers, and the distance between them.
For the tunnel barriers inducing a quasi-discrete spectrum, in the case of weak coupling to the Fano state there is a single emergent resonance located near the Fano-state energy. For strong coupling, there may emerge several resonances with the heights of the order of unity. For an asymmetric setup with strong non-equal barriers, the transmission coefficient of the double-barrier structure is suppressed even at quantization levels of the double-barrier structure, but an almost perfect transmission may occur through the structure with a side-attached state. This phenomenon of the transmission enhancement by the Fano antiresonance can be employed for the ``transport filtering'' of monochromatic waves in electronic, optical, or microwave applications.

\begin{acknowledgments}
    We are grateful to N.S. Averkiev, S. Klyatskaya, M. Ruben, and W. Wernsdorfer for useful discussions.
    We acknowledge collaboration with P. Shmakov at the early stage of this work.
	The research was supported by the Russian Science Foundation
	 (Grant No. 17-12-01182), the Russian
	Foundation for Basic Research (Grant No. 18-02-01016), and by the Foundation for the advancement of 
	theoretical physics and mathematics
	``BASIS'' (I.V.K.).
\end{acknowledgments}

\vspace{0.3cm}

\begin{appendix}

\section{Scattering-matrix approach}

In this Appendix, we calculate the transmission amplitude across a double-barrier structure with a Fano center using a scattering-matrix approach. The transmission amplitude is obtained by a direct summation of the plain waves reflected inside the structure. 
The barriers are located at $x = \pm L$ and the Fano center is side-attached at $x = d$. 
Consider first a combined scatterer fromed by the left barrier and the Fano state. 
For this scatterer, we obtain the transmission amplitude $t_\text{LF}$ and the reflection amplitude $r_\text{FL}$ describing the plane wave to the right of the Fano center:
\begin{gather}
	t_\text{LF} = \frac{t_L t_F \e^{i k (L + d)}}{1 - r_L r_F \e^{i 2 k (L + d)}}, 
\label{tLF}	
	\\[0.2cm]
	r_\text{FL} = \frac{r_F- r_L r_F^2 \e^{i 2 k (L + d)} + r_L t_F^2 \e^{i 2 k (L + d)}}
	{1 - r_L r_F \e^{i 2 k (L + d)}}.
	\label{rFL}
\end{gather}

\begin{widetext}
Now, we can obtain the transmission amplitude across the whole (Left barrier--Fano--Right barrier) structure by replacing the left barrier and the Fano center with the effective complex scatterer characterized by the scattering amplitudes (\ref{tLF}) and (\ref{rFL}):
\begin{gather}
	t_\text{LFR} = \frac{t_\text{LF} t_R \e^{i k (L- d)}}{1- r_L r_\text{FL} \e^{i 2 k (L- d)}} = 
	\frac{t_{L} t_F t_R \e^{i 2 k L}}{1- r_R r_F \e^{i 2 k (L+d)} - r_R r_F \e^{i 2 k (L - d)} - r_R r_L (t_F^2 - r _F^2) \e^{i 4 k L}}.
	\label{tlFr}
\end{gather}
Using the relation between the transmission and reflection amplitudes for the point-like scatterer, 
$t_F - r_F = 1$, we write
\begin{gather}
t_\text{LFR} = \frac{t_L t_F t_R \e^{i 2 k L}}{1 - r_L r_F \e^{i 2 k (L+d)} - r_R r_F \e^{i 2 k (L - d)} 
- r_L r_R \e^{i 4 k L} - 2 r_L r_R r_F \e^{i 4 k L}}.
\label{tLFR-res-0}
\end{gather}
After straightforward algebra, we arrive at Eq. (\ref{tLFR-result}) of the main text.

We can simplify Eq.~(\ref{tLFR-result}) for strong barriers, $\mathcal{T}_L,\mathcal{T}_R\ll 1$,  and far from the quantization level of the double-barrier structure $|k-k_n|L \gg \mathcal{T}_L+\mathcal{T}_R$ [see Eq.~(\ref{transmission-BB-asym})]. To gain intuition about the structure of $t_\text{LFR}$, we
replace $\e^{i 4 k L} \to -1$ in Eq.~\eqref{tLFR-result}, assume $|t_{R,L}|\ll 1$, and use the relation $r_{R,L}=1-t_{R,L}$ valid for $\delta$-barriers. This results in a simplified expression: 
(we do not give the explicit expression for the broadening $\gamma$ here, as it is not important for finding the position of the resonance):
\begin{equation}
t_\text{LFR}\simeq \frac{(E_k - \varepsilon_0)t_\text{BB}}{E_k-\varepsilon_0+\Gamma_0 \cos(2 kd) + i\gamma},
\quad \gamma\ll \Gamma_0.
\end{equation}
This yields the relation
\begin{equation}
E_k\simeq \varepsilon_0-\Gamma_0 \cos(2 kd)
\label{emergent-spectrum}
\end{equation}
for the energy of the emergent resonances. Depending on the relation between $\Gamma_0$ and the level spacing of the double-barrier structure $\Delta$, this relation yield one (small $\Gamma_0$) or many solutions.

Returning to Eq. (\ref{tLFR-result}) away from the resonances in $\mathcal{T}_\text{BB}$,
we write
\begin{gather}
\mathcal{T}_\text{BFB} \approx \frac{(E_k - \varepsilon_0)^2 \mathcal{T}_\text{BB}}
{\left\{E_k - \varepsilon_0 -\Gamma_0 \frac{2 \sin[k(L-d)] \sin[k(L+d)]}{\sin(2 k L)} \right\}^2 
+ \Gamma_0^2 
\left\{ \frac{\mathcal{T}_L \sin^2[k(L-d)] + \mathcal{T}_R \sin^2[k(L+d)]}{2 \sin^2(2 k L)} \right\}^2}.
\label{TBFB-strong-asym}
\end{gather}
Here we have neglected, for simplicity, the phases of the reflection amplitudes of the barriers (they can be incorporated into the redefinition of $L$ and $d$), assuming the delta-barriers. 
The transmission coefficient at the emergent resonance  reads:
\begin{gather}
	\mathcal{T}_\text{BFB,res} \approx 
	\left.\mathcal{T}_\text{BB} \left\{ \frac{4 \sin [k (L-d)] \sin[k(L+d)] 
	\sin (2kL)}{\mathcal{T}_L  \sin^2[k (L - d)] + \mathcal{T}_R  \sin^2[k(L+d)]} \right\}^2
	\right|_{k=k_\text{res}},
	\label{TBFB-res-asym}
\end{gather}
where $k_\text{res}$ corresponds to a maximum of $\mathcal{T}_\text{BFB}$ in Eq.~(\ref{TBFB-strong-asym}). This representation of 
$\mathcal{T}_\text{BFB,res}$ leads to Eq.~(\ref{TBFBres-general}), and allows us to estimate the conditions 
(\ref{res-condition}) and (\ref{res-accuracy}) for the emergence of the resonance presented in Sec.~\ref{Sec:Resonance} of the main text.

\end{widetext}

Finally, returning to Eq.~(\ref{tlFr}), we set $r_R=0$, $t_R=1$, $r_L=r_B$, and $t_L=t_B$ to get the transmission amplitude for the single-barrier setup discussed in Sec.~\ref{sec:BF}:
\begin{gather}
t_\text{BF} = \frac{t_B t_F \e^{i 2 k L}}{1 - r_B r_F \e^{i 2 k (L+d)}}.
\end{gather}
With $t_F$ and $r_F$ from Eqs.~(\ref{trFano}) and (\ref{refFano}), this becomes
\begin{align}
\label{t_BF}
t_\text{BF}=\frac{E_k-\varepsilon_0}
{E_k-\varepsilon_0-\frac{2\pi i m}{k}|V|^2\left(1+r_B e^{4ik L}\right)}\, t_B.
\end{align}

\end{appendix}


\begin{thebibliography}{18}
\bibitem{footnote}
We would like to mention that it would be interesting to study the emergent resonances in the presence of electron-electron and electron-phonon interactions. For this purpose, the formalism for the calculation of transmission through a double-barrier structure with the Fano state developed in Ref. \cite{Klier-PhD} on the basis of a Green's function approach, generalizing the one described in the main text, might be useful.
\end{thebibliography}
\end{document}